\documentclass{elsart}

\usepackage{amssymb}
\usepackage[fleqn]{amsmath}
\usepackage{graphicx}
\usepackage{subfigure}
\usepackage[active]{srcltx}
\usepackage{hhline, portland}
\usepackage[official,right]{eurosym}
\usepackage{multicol}
\usepackage{rotating}
\usepackage{array}
\usepackage{multirow}

\usepackage{lscape}
\usepackage{amsfonts}

\begin{document}

\begin{frontmatter}
\title{True and Apparent Scaling: The Proximity of the Markov-Switching Multifractal Model to Long-Range Dependence}
\author[Kiel,Canberra]{Ruipeng Liu},
\author[Canberra]{T. Di Matteo},
\author[Kiel]{Thomas Lux}
\address[Kiel]{Department of Economics, University of Kiel, 24118 Kiel, Germany}
\address[Canberra]{Department of Applied Mathematics, Research School of Physical Sciences and Engineering, The Australian National University, 0200 Canberra, Australia}
\date{}

\begin{abstract}
In this paper, we consider daily financial data of a collection of
different stock market indices, exchange rates, and interest
rates, and we analyze their multi-scaling properties by estimating
a simple specification of the Markov-switching multifractal model
(MSM). In order to see how well the estimated models capture the
temporal dependence of the data, we estimate and compare the
scaling exponents $H(q)$ (for $q = 1, 2$) for both empirical data
and simulated data of the estimated MSM models. In most cases the
multifractal model appears to generate `apparent' long memory in
agreement with the empirical scaling laws.

\end{abstract}
\end{frontmatter}

\section{Introduction}
The scaling concept has its origin in physics but it is
increasingly applied outside its traditional domain. In the
literature (\cite{muller90,Dacorogna01a,Dacorogna01b}) different
methods have been proposed and developed in order to study the
multi-scaling properties of financial time series. For more
details on scaling analysis see \cite{DiMatteo06}.

Going beyond the phenomenological scaling analysis, the
multifractal model of asset returns (MMAR) introduced by
Mandelbrot et. al \cite{Mandelbrot97} provides a theoretical
framework that allows to replicate many of the scaling properties
of financial data. While the practical applicability of MMAR
suffered from its combinatorial nature and its non-stationarity,
these drawbacks have been overcome by the introduction of
iterative multifractal models (Poisson MF or Markov-switching
multifractal model (MSM) \cite{Calvet01,Calvet04,Lux05}) which
preserves the hierarchical, multiplicative structure of the
earlier MMAR, but is of much more `well-behaved' nature concerning
its asymptotic statistical properties. The attractiveness of MF
models lies in their ability to mimic the stylized facts of
financial markets such as outliers, volatility clustering, and
asymptotic power-law behavior of autocovariance functions
(long-term dependence). In contrast to other volatility models
with long-term dependence \cite{Baillie}, MSM models allow for
multi-scaling rather than uni-scaling with varying decay exponents
for all powers of absolute values of returns. One may note,
however, that due to the Markovian nature, the scaling of the
Markov-Switching MF model only holds over a limited range of time
increments depending on the number of hierarchical components and
this `apparent' power-law ends with a cross-over to an exponential
cut-off.

With this proximity to true multi-scaling, it seems worthwhile to
explore how well the MSM model could reproduce the empirical
scaling behaviour of financial data. To this end, we estimate the
parameters of a simple specification of the MSM model for various
financial data and we assess its ability to replicate empirical
scaling behaviour by also computing $H(q)$ by means of the
generalized Hurst exponent approach
(\cite{DiMatteo06,DiMatteo03,DiMatteo05}) and $H$ by means of the
modified R/S method \cite{Lo91} for the same data sets. We then
proceed by comparing the scaling exponents for empirical data and
simulated time series based on our estimated MSM models. As it
turns out, the MSM model with a sufficient number of volatility
components generates pseudo-empirical scaling laws in good overall
agreement with empirical results.

The structure of the paper is as follows: In Section 2 we
introduce the multifractal model, the Generalized Hurst exponent
(GHE) and the modified R/S approaches. Section 3 reports the
empirical and simulation-based results. Concluding remarks and
perspectives are given in Section 4.

\section{Methodology}

\subsection{Markov-switching multifractal model}

In this section, we shortly review the building blocks of the
Markov-switching multifractal process (MSM). Returns are modeled
as \cite{Calvet04,Lux05}:
\begin{equation}\label{EQ2}
    r_{t}=\sigma_{t}\cdot u_{t}
\end{equation}

with innovations $u_{t}$ drawn from a standard Normal distribution
$N(0,1)$ and instantaneous volatility being determined by the
product of $k$ volatility components or multipliers $M_{t}^{(1)}$,
$M_{t}^{(2)}$ ..., $M_{t}^{(k)}$ and a constant scale factor
$\sigma$:
\begin{equation}\label{EQ3}
    \sigma^2_{t}=\sigma^2\prod^k_{i=1} M_{t}^{(i)},
\end{equation}

In this paper we choose, for the distribution of volatility
components, the binomial distribution: $M_t^{(i)} \sim [m_0,
\hspace{0.2cm} 2-m_0]$ with $1 \leq m_0 < 2$. Each volatility
component is renewed at time $t$ with probability $\gamma_i$
depending on its rank within the hierarchy of multipliers and it
remains unchanged with probability $1-\gamma_i$. The transition
probabilities are specified by Calvet and Fisher \cite{Calvet04}
as:
\begin{equation}\label{EQ4}
    \gamma_{i}=1-(1-\gamma_{k})^{(b^{i-k})} \hspace{2cm} i = 1, \ldots k,
\end{equation}

with parameters $\gamma_{k}\in[0,1]$ and $b\in(1,\infty)$.
Different specifications of  Eq. (\ref{EQ4}) can be arbitrarily
imposed (cf. \cite{Lux05} and its earlier versions). By fixing $b
= 2$ and $\gamma_k = 0.5$, we arrive a relatively parsimonious
specification:
\begin{equation}\label{EQ5}
    \gamma_{i}=1-(1-\gamma_k)^{(2^{i-k})} \hspace{2cm} i = 1, \ldots
    k.
\end{equation}

This specification implies that replacement happens with
probability of one half at the highest cascade level. Various
approaches have been employed to estimate multifractal models. The
parameters of the combinatorial MMAR have been estimated via an
adaptation of the scaling estimator and Legendre transformation
approach from statistical physics \cite{Calvet02}. However, this
approach has been shown to yield very unreliable results
\cite{Lux04a}.  A broad range of more rigorous estimation methods
have been developed for the MSM model. Calvet and Fisher (2001)
(\cite{Calvet01}) propose maximum likelihood estimation while Lux
(\cite{Lux05}) proposes a Generalized Method of Moments (GMM)
approach, which can be applied not only to discrete but also to
continuous distributions of the volatility components. In this
paper, GMM is used to estimate the two MSM model parameters in Eq.
(\ref{EQ3}), namely: $\hat{\sigma}$ and $\hat{m_0}$.

\subsection{Estimation of scaling exponents}

Our analysis of the scaling behaviour of both empirical and
simulated data uses two refined methods for estimating the
time-honored Hurst coefficient: the estimation of generalized
Hurst exponents from the structure function of various moments
\cite{DiMatteo06} and Lo's modified R/S analysis that allows to
correct for short-range dependence in the temporal evolution of
the range \cite{Lo91}.

\subsubsection{Generalized Hurst exponent approach}

The generalized Hurst exponent (GHE) method extends the
traditional scaling exponent methodology, and this approach
provides a natural, unbiased, statistically and computationally
efficient estimator able to capture very well the scaling features
of financial fluctuations (\cite{DiMatteo03,DiMatteo05}). It is
essentially a tool to study directly the scaling properties of the
data via the $q$th order moments of the distribution of the
increments. The $q$th order moments appear to be less sensitive to
the outliers than maxima/minima and different exponents $q$ are
associated with different characterizations of the multi-scaling
behaviour of the signal $X(t)$.

We consider the $q$-order moment of the distribution of the
increments (with $t = v, 2v, . . . ,T)$ of a time series $X(t)$:
\begin{equation}\label{K_q}
K_q(\tau) = \frac{\langle \mid X(t+\tau) - X(t)
\mid^q\rangle}{\langle \mid X(t)\mid^q\rangle},
\end{equation}

where the time interval $\tau$ varies between $v = 1$ day and
$\tau_{max}$ days. The generalized Hurst exponent $H(q)$ is then
defined from the scaling behavior of $K_q(\tau)$, which can be
assumed to follow the relation:
\begin{equation}\label{K_qtao}
K_q(\tau) \sim {\left(\frac{\tau}{v}\right )}^{qH(q)}.
\end{equation}

Within this framework, for $q = 1$, $H(1)$ describes the scaling
behavior of the absolute values of the increments; for $q = 2$,
$H(2)$ is associated with the scaling of the autocorrelation
function.

\subsubsection{Lo's modified R/S analysis}

Lo's modified R/S analysis uses the range of a time series as its
starting point: Formally, the range $R$ of a time series
$\{X_t\}$, $t = 1, \ldots, T$ is defined as:
\begin{equation}\label{Lo1}
R_T = \max_{1\leq t \leq T} \sum_{t=1}^T(X_t - \bar{X}) -
\min_{1\leq t \leq T} \sum_{t=1}^T(X_t - \bar{X}).
\end{equation}
Here, $\bar{X}$ is the standard estimate of the mean. Usually the
range is rescaled by the sample standard deviation ($S$), yielding
the famous R/S statistic. Though this approach found wide
applications in diverse fields, it turned out that no asymptotic
distribution theory could be derived for H itself. Hence, no
explicit hypothesis testing can be performed and the significance
of point estimates $H > 0.5$ or $H < 0.5$ rests on subjective
assessment. Luckily, the asymptotic distribution of the rescaled
range itself under a composite null hypothesis excluding
long-memory could be established by Lo (1991) \cite{Lo91}. Using
this distribution function and the critical values reported in his
paper, one can test for the significance of apparent traces of
long memory as indicated by $H \neq 0.5$. However, Lo also showed
that the distributional properties of the rescaled range are
affected by the presence of short memory and he devised a modified
rescaled range $Q_\tau$ which adjusts for possible short memory
effects by applying the Newey-West heteroscedasticity and
autocorrelation consistent estimator in place of the sample
standard deviation $S$:
\begin{eqnarray}\label{Lo2}
Q_\tau &=& \frac{1}{S_\tau}\left[\max_{1\leq t \leq T}
\sum_{t=1}^T(X_t - \bar{X}) - \min_{1\leq t \leq T}
\sum_{t=1}^T(X_t - \bar{X})\right],\\
\nonumber\\
S_{\tau}^2 &=& S^2 + \frac{2}{T}\sum_{j=1}^\tau
\omega_j(\tau)\left\{\sum_{i=j+1}^T(X_i - \bar{X})(X_{i-j} -
\bar{X})\right\},\nonumber\\
\omega_j(\tau) &=& 1 - \frac{j}{\tau + 1}.\nonumber
\end{eqnarray}
Under the null of no long term memory the distribution of the
random variable $V_T = T^{-0.5}Q_\tau$ converges to that of the
range of a so-called Brownian bridge. Critical values of this
distribution are tabulated in Lo (1991, Table II).

\section{Results}
In this paper, we consider daily data for a collection of stock
exchange indices: the Dow Jones Composite 65 Average Index ($Dow$)
and $NIKKEI$ 225 Average Index ($Nik$) over the time period from
January 1969 to October 2004, foreign exchange rates: British
Pound to US Dollar ($UK$), and Australian Dollar to US Dollar
($AU$) over the period from March 1973 to February 2004, and U.S.
1 year and 2 years treasury constant maturity bond rates ($TB1$
and $TB2$, respectively) in the period from June 1976 to October
2004. The daily prices are denoted as $p_t$, and returns are
calculated as $r_t = \ln(p_t) -\ln(p_{t-1})$ for stock indices and
foreign exchange rates and as $r_t = p_t - p_{t-1}$ for $TB1$ and
$TB2$.

We estimate  the MSM model parameters introduced in Section 2 with
a binomial distribution of volatility components, that is
$M_t^{(\cdot)} \sim [m_0, \hspace{0.2cm} 2-m_0]$ and $1 \leq m_0 <
2$ in Eq \ref{EQ3}.  This estimation is repeated for various
hypothetical numbers of cascade levels ($k = 5, 10, 15, 20$).
Table \ref{mf_estK} presents these results for parameters
$\hat{m_0}$ and $\hat{\sigma}$.\footnote{Note that the data have
been standardized by dividing the sample standard deviation which
explains the proximity of the scale parameter estimates to 1.} Our
estimation is based on the GMM approach proposed by Lux
\cite{Lux05} using the same analytical moments as in his paper.
The numbers within the parentheses are the standard errors. We
observe that the results for $k>10$ are almost identical. In fact,
analytical moment conditions in Lux \cite{Lux05} show that higher
cascade levels make a smaller and smaller contribution to the
moments so that their numerical values would stay almost constant.
If one monitors the development of estimated parameters with
increasing $k$, one finds strong variations initially with a
pronounced decrease of the estimates which become slower and
slower until, eventually a constant value is reached somewhere
around $k=10$ depending on individual time series. Based on the
estimated parameters, we proceed with an analysis of simulated
data from the pertinent MSM models.

We first calculate the GHE for the empirical time series as well
as for $100$ simulated time series of each set of estimated
parameters for $q = 1$ and $q = 2$. The values of the GHE are
averages computed from a set of values corresponding to different
$\tau_{max}$ (between 5 and 19 days). The stochastic variable
$X(t)$ in Eq. (\ref{K_q}) is the absolute value of returns, $X(t)
= |r_t|$. The second and seventh columns in Table \ref{compare}
report the empirical GHEs, and values in the other columns are the
mean values over the corresponding 100 simulations for different
$k$ values: $5, 10, 15, 20$, with errors given by their standard
deviations. Boldface numbers are those cases which fail to reject
the null hypothesis  that the mean of the simulation-based
Generalized Hurst exponent values equals the empirical Generalized
Hurst exponent at the $5\%$ level. We find that the exponents from
the simulated time series vary across different cascade levels
$k$. In particular, we observe considerable jumps from $k=5$ to
$k=10$ for these values. In particular for the stock market
indices, we find coincidence between the empirical series and
simulation results for the scaling exponents $H(2)$ for $Dow$ and
$H(1)$ for $Nik$ when $k = 5$. For the exchange rate data, we
observe the simulations successfully replicate the empirical
measurements of $AU$ for $H(1)$ when $k = 10, 15, 20$ and $H(2)$
when $k = 5$; In the case of U.S. Bond rates, we find a good
agreement for $H(1)$ when $k = 5$ and for all $k$ for $TB1$, and
$H(2)$ for $TB2$ when $k = 5$. Apparently, both the empirical data
and the simulated MSM models are characterized by estimates of
$H(1)$ and $H(2)$ much larger than 0.5 which are indicative of
long-term dependence. While the empirical numbers are in nice
agreement with previous literature,
 it is interesting to note that simulated data with $k \geq 10$
have a tendency towards even higher estimated Hurst coefficients
than found in the pertinent empirical records.\footnote{We have
checked if the generalized Hurst exponents approach is biased by
computing $H(1)$ and $H(2)$ for random values generated by
different random generators \cite{DiMatteo05} with $T = 9372$ data
points. We have found that $H(1) = 0.4999 \pm 0.009$ and $H(2) =
0.4995 \pm 0.008$.} Since we know that the MSM model only has
pre-asymptotic scaling, these results underscore that with a high
enough number of volatility cascades, it would be hard to
distinguish the MSM model from a `true' long memory process.

We have also performed calculations using the modified Rescaled
range (R/S) analysis introduced by Lo
\cite{Lo91,Mills93,Huang95,Brook95,Lux96,Crato96,Williger99},\footnote{We
also did a Monte Carlo study with 1000 simulated  random time
series in order to assess the bias of the pertinent estimates of
$H$: for random numbers with sample size $T=9372$ (comparable to
our empirical records) we obtained a slight negative bias:
$H=0.463\pm 0.024$.} whose results are reported in Tables
\ref{Lo'sV} to \ref{Lo'sH}. Table \ref{Lo'sV} presents Lo's test
statistics for both empirical and 1000 simulated time series for
different values of $k$ and for different truncation lags $\tau =
0, 5, 10, 25, 50, 100$.\footnote{For $\tau = 0$ we have the
classical R/S approach.} We find that the values are varying with
different truncation lags, and more specifically, that they are
monotonically decreasing for both the empirical and
simulation-based statistics. Table \ref{rejections} reports the
number of rejections of the null hypothesis of short-range
dependence based on $95\%$ and $99\%$ confidence levels. The
rejection numbers for each single $k$ are decreasing as the
truncation lag $\tau$ increases, but the proportion of rejections
remains relatively high for higher cascade levels, $k = 10, 15,
20$.  The corresponding Hurst exponents are given in Table
\ref{Lo'sH}. The empirical values of $H$ are decreasing when
$\tau$ increases. A similar behaviour is observed for the
simulation-based $H$ for given values of $k$. We also observe that
the Hurst exponent values are increasing with increasing cascade
level $k$ for given $\tau$. Boldface numbers are those cases which
fail to reject the null hypothesis that the mean of the
simulation-based Hurst exponent equals the empirical Hurst
exponent at the $5\%$ level. There are significant jumps between
the values for $k=5$ and $k=10$ as reported in previous tables.

Overall, the following results stand out: (1) There seems to be a
good overall agreement between the empirical and simulated data
for practically all series for levels $k \geq 10$, while with a
smaller number of volatility components ($k = 5$) the simulated
MSM models have typically smaller estimated $H$s than the
corresponding empirical data, (2) the modified R/S approach would
quite reliably reject the null of long memory for $k = 5$, but in
most cases it would be unable to do so for higher numbers of
volatility components, even if we allow for large truncation lags
up to $\tau = 100$. Results are also much more uniform than with
the generalized Hurst technique which had left us with a rather
mixed picture of coincidence of Hurst coefficients of empirical
and simulated data. The fact, that according to Table \ref{Lo'sH},
MSM model with 15 or more volatility components did always produce
`apparent' scaling in agreement with that of empirical data, is
particular encouragingly. It contrasts with the findings reported
in \cite{Crato96} on apparent scaling of estimated GARCH models
whose estimated exponents did not agree with the empirical ones.

\section{Concluding Remarks}

We have calculated the scaling exponents of simulated data based
on estimates of the Markov-switching multifractal (MSM) model.
Comparing the generalized Hurst exponent values as well as Lo's
Hurst exponent statistics of both empirical and simulated data,
our study shows that the MSM model captures quite satisfactorily
the multi-scaling properties of absolute values of returns for
specifications with a sufficiently large number of volatility
components. Subsequent work will explore whether this encouraging
coincidence of the scaling statistics for the empirical and
synthetic data also holds for other candidate distributions of
volatility components and alternative specifications of the
transition probabilities.

\textbf{Acknowledgments}

T. Di Matteo acknowledges the partial support by ARC Discovery
Projects: DP03440044 (2003) and DP0558183 (2005), COST P10
``Physics of Risk" project and M.I.U.R.-F.I.S.R. Project
``Ultra-high frequency dynamics of financial markets", T. Lux
acknowledges financial support by the European Commission under
STREP contract No. 516446.


\begin{table}[!h]
\caption{GMM estimates of MSM model for different values of
k.}\label{mf_estK} {\small
\begin{tabular}{|c|cc|cc|cc|cc|}
\hline &\multicolumn{2}{|c|}{k = 5}&\multicolumn{2}{|c|}{k =
10}&\multicolumn{2}{|c|}{k =
15}&\multicolumn{2}{|c|}{k = 20}\\
\hline
&  $\hat{m}_0$ & $\hat{\sigma}$&  $\hat{m}_0$ & $\hat{\sigma}$ & $\hat{m}_0$ & $\hat{\sigma}$ & $\hat{m}_0$ & $\hat{\sigma}$ \\
\hline
$Dow$ &1.498& 0.983 & 1.484  & 0.983  & 1.485  & 0.983 & 1.487  &  0.983  \\
      &(0.025)& (0.052)  &(0.026)  &(0.044)  &(0.026)&(0.042)&(0.027)  &(0.044)\\
$Nik$  &1.641& 0.991&1.634  &0.991  & 1.635  &0.991 & 1.636  & 0.991  \\
      &(0.017)& (0.036) &(0.013)&(0.028)&(0.017) &(0.036)&(0.017)&(0.037)\\
$UK$  &1.415& 1.053 &1.382  & 1.057& 1.381  & 1.056  & 1.381 & 1.058 \\
       &(0.033)&(0.026) &(0.029) &(0.027)&(0.036)&(0.027)&(0.038)&(0.026)\\
$AU$   &1.487& 1.011&1.458  &1.013   &1.457  &1.014& 1.458& 1.014 \\
       &(0.034)&(0.066) &(0.034) &(0.061)&(0.034)&(0.066)&( 0.034)&(0.065)\\
$TB1$  &1.627& 1.041&1.607  &1.064  &1.607 &1.064  &1.606  & 1.067  \\
       &(0.021)&(0.032) &(0.025) &(0.024)&(0.028)&(0.024)&(0.025)&(0.024)\\
$TB2$  &1.703& 1.040&1.679  &1.068  &1.678  &1.079  &1.678 & 1.079 \\
      &(0.015)&(0.036)  &(0.014) &(0.029)&(0.015)&(0.032)&(0.015)&(0.034)\\
\hline
\end{tabular}\\[0.5ex]
}\footnotesize{Note: All data have been standardized before
estimation.}
\end{table}


\begin{table}[!h]
\caption{H(1) and H(2) for the empirical and simulated
data.}\label{compare}{\small
\begin{tabular}{|c|ccccc|ccccc|}
\hline &\multicolumn{5}{|c|}{$H(1)$}&\multicolumn{5}{|c|}{$H(2)$}\\
\hline
& $Emp$ & $sim1$ & $sim2$ & $sim3$ & $sim4$ &$Emp$ & $sim1$ & $sim2$ & $sim3$&$sim4$ \\
\hline
$Dow$    & 0.684   &0.747   &0.849   &0.868   &0.868                             & 0.709  & \textbf{0.705} &0.797 &0.813   &0.812\\
         &(0.034)   &(0.008)   &(0.015)   &(0.021)   &(0.024)                    &(0.027)&(0.009) &(0.015)   &(0.019)   &(0.022) \\
$Nik$   &0.788   & \textbf{0.801}   &0.894   &0.908  & 0.908                     &0.753   & {0.736} &0.815 &0.824   &0.824\\
         &(0.023)   &(0.008)   &(0.013)   &(0.019)   &(0.028)                    &(0.021)&(0.008) &(0.013)  &(0.018)  &(0.024)\\
 $UK$    &0.749   &0.709   &0.799   &0.825   &0.821                              &0.735    &0.678 & {0.764} &0.785   &0.783\\
         &(0.023)   &(0.010)   &(0.018)   &(0.025)   &(0.026)                    &(0.026)&(0.010) &(0.016)   &(0.021)   &(0.022)\\
 $AU$    &0.827   &0.746   &\textbf{0.837}  &\textbf{0.860}   &\textbf{0.857}    &0.722   & \textbf{0.705} &0.790 &0.808   &0.808\\
         &(0.017)   &(0.009)   &(0.016)   &(0.022)   &(0.021)                    &(0.024) &(0.009) &(0.015)   &(0.018)   &(0.018)\\
 $TB1$   &0.853   & \textbf{0.856}   &0.909   &0.915   &0.911                    &0.814  & \textbf{0.783} &\textbf{0.826} &\textbf{0.832}   &\textbf{0.829}\\
         &(0.022)   &(0.035)   &(0.023)   &(0.026)   &(0.026)                    &(0.027)&(0.028) &(0.020)   &(0.020)  &(0.020)\\
 $TB2$   &0.791   &0.866   &0.920   &0.924   &0.919                              &0.778   & \textbf{0.781} &0.823 &0.827  &{0.822}\\
         &(0.025)   &(0.029)   &(0.021)   &(0.022)   &(0.026)                    &(0.029)&(0.022) &(0.017)   &(0.022)   &(0.023)\\

 \hline
\end{tabular}\\[0.5ex]
}\footnotesize{Note: $Emp$ refers to the empirical exponent
values, sim1, sim2, sim3 and sim4 are the corresponding exponent
values based on the simulated data for $k=5$, $k=10$, $k=15$ and
$k=20$ respectively. The stochastic variable $X_t$ is defined as
$|r_t|$. Bold numbers show those cases for which we cannot reject
identity of the Hurst coefficients obtained for empirical and
simulated data, i.e. the empirical exponents fall into the range
between the 2.5 to 97.5 percent quantile of the simulated data.}
\end{table}


\begin{landscape}
\begin{table}[!h]
\caption{Lo's R/S statistic for the empirical and simulated
data.}\label{Lo'sV}{\scriptsize
\begin{tabular}{|c|ccccc|ccccc|ccccc|}
\hline &\multicolumn{5}{|c|}{$\tau=0$}&\multicolumn{5}{|c|}{$\tau=5$}&\multicolumn{5}{|c|}{$\tau=10$}\\
\hline
& $Emp$ & $k=5$ & $k=10$ & $k=15$ & $k=20$ &$Emp$ & $k=5$ & $k=10$ & $k=15$ & $k=20$&$Emp$ & $k=5$ & $k=10$ & $k=15$ & $k=20$  \\
\hline
&&&&&  &&&&& &&&&&\\
$Dow$    &3.005 &1.712 &5.079 &6.640 &6.704         &2.661 &1.481 &4.060 &5.211 &5.263       & 2.427 &1.376 &3.574 &4.537 & 4.582                    \\
         &&(0.381)&(1.300)&(1.769)&(1.839)              &&(0.329)&(1.017)&(1.333)&(1.387)                &&(0.305)&(0.884)&(1.133)&(1.179)\\
$Nik$    & 7.698&1.840 &4.898&6.154 &6.152             &6.509 &1.540&3.817 &4.747 &4.742        &5.836 &1.416 &3.343 &4.132 &4.133                     \\
          &&(0.425)&(1.195)&(1.520)&(1.584)                  &&( 0.355)&(0.918)&(1.147)&(1.193)                &&(0.325)&(0.798)&(0.984)&(1.023)\\
 $UK$   &6.821 &1.544 &4.599 &6.047 &6.175              &5.912  &1.370 &3.815 &4.918 &5.008             &5.333 &1.286 &3.405 &4.337 &4.408                      \\
          &&(0.350)&(1.200)&(1.748)&(1.848)         &&(0.310)&(0.972)&(1.352)&(1.417)             &&(0.290)&(0.854)&(1.157)&(1.207)\\
$AU$   &7.698 &1.687 &4.962 &6.348 & 6.434               & 6.731 &1.463 &4.001 &5.024 &5.090              &6.103 &1.361 &3.531 &4.387 &4.443                     \\
          &&(0.386)&(1.257)&(1.742)&(1.790)           &&(0.333)&(0.989)&(1.315)&(1.352)            &&(0.309)&(0.861)&(1.117)&(1.149)\\
 $TB1$  &8.845 &1.826 &4.644 &5.915 &6.041            &7.109 &1.524 &3.629 &4.564 & 4.582              &6.110 &1.400 &3.184 &4.415 &4.530                     \\
          &&(0.398)&(1.141)&(1.425)&(1.380)           &&(0.330)&(0.875)&(1.074)&(1.040)              &&(0.302)&(0.759)&(0.921)&(0.891)\\
 $TB2$   &7.295 &1.855 &4.347 &5.853 & 5.907       &6.083 &1.531 &3.391 &4.207 &4.349           &5.330 &1.404 &2.985 &4.025 &4.158                     \\
          &&(0.413)&(1.031)&(1.215)&(1.227)           &&(0.339)&(0.795)&(0.928)&(0.930)                      &&(0.310)&(0.694)&(0.804)&(0.803)\\
&&&&&  &&&&& &&&&&\\
\hline

&\multicolumn{5}{|c|}{$\tau=25$}&\multicolumn{5}{|c|}{$\tau=50$}&\multicolumn{5}{|c|}{$\tau=100$}\\
\hline
& $Emp$ & $k=5$ & $k=10$ & $k=15$ & $k=20$ &$Emp$ & $k=5$ & $k=10$ & $k=15$ & $k=20$&$Emp$ & $k=5$ & $k=10$ & $k=15$ & $k=20$  \\
\hline
&&&&&  &&&&& &&&&&\\
$Dow$    &2.042 &1.237 &2.877 &3.580 & 3.616             &1.736 &1.153 &2.385 &2.909 & 2.941      &1.464 &1.098 &1.965 &2.338 & 2.366                    \\
         &&(0.272)&(0.694)&(0.857)&(0.893)          &&(0.250)&(0.560)&(0.668)&(0.696)             &&(0.233)&(0.443)&(0.508)&(0.530)\\
$Nik$    &4.760 &1.260 &2.692 &3.285 &3.279           &3.941 &1.169 &2.246 &2.701 &2.698       &3.220 &1.113 &1.868 &2.204 & 2.203                   \\
          &&(0.286)&(0.631)&(0.761)&(0.788)                  &&(0.263)&(0.514)&(0.604)&(0.623)         &&(0.245)&(0.412)&(0.468)&(0.482)\\
 $UK$    &4.348 &1.170 &2.782 &3.469 & 3.515               &3.575 &1.099&2.322 &2.837 &2.868        &2.871 &1.053 &1.922 &2.289 &2.306                     \\
          &&(0.262)&(0.678)&(0.876)&(0.909)        &&(0.244)&(0.549)&(0.680)&(0.702)           &&(0.228)&(0.434)&(0.513)&(0.528)\\
$AU$    &5.035 &1.224 &2.848 &3.474 & 3.516                          &4.130 &1.142 &2.362 &2.830 &2.861              &3.281 &1.089 &1.947 &2.280 &2.302                     \\
          &&(0.275)&(0.676)&(0.842)&(0.866)                  &&(0.252)&(0.544)&(0.654)&(0.672)                      &&(0.232)&(0.429)&(0.496)&(0.508)\\
 $TB1$   &4.580 &1.245 &2.571 &2.961 &2.971           &3.514 &1.156 &2.148 &2.442 &2.449             &2.649 &1.101 &1.790 &2.004 &2.006                     \\
          &&(0.265)&(0.598)&(0.711)&(0.685)                  &&(0.242)&(0.484)&(0.564)&(0.542)                      &&(0.223)&(0.384)&(0.440)&(0.417)\\
 $TB2$   &4.129 &1.249 &2.432 &2.762 & 2.786             &3.250 &1.162 &2.052 &2.305 & 2.320              &2.502 &1.109 &1.731 &1.915 & 1.921                   \\
          &&(0.272)&(0.554)&(0.632)&(0.630)                  &&(0.249)&(0.456)&(0.511)&(0.507)                      &&(0.230)&(0.369)&(0.403)&(0.398)\\
&&&&&  &&&&& &&&&&\\
 \hline
\end{tabular}\\[0.5ex]
 \scriptsize{Note: $Emp$ stands for the empirical Lo's statistic, $k=5$, $k=10$, $k=15$ and $k=20$ refer to the mean and standard
deviation of Lo's statistics based \\ on the corresponding 1000
simulated time series with pertinent $k$.}}
\end{table}

\end{landscape}


\begin{landscape}
\begin{table}[!h]
\caption{Number of rejections for Lo's R/S statistic
test.}\label{rejections}{\scriptsize
\begin{tabular}{|c|cccccccc|cccccccc|cccccccc|}
\hline
&\multicolumn{8}{|c|}{$\tau=0$}&\multicolumn{8}{|c|}{$\tau=5$}&\multicolumn{8}{|c|}{$\tau=10$}\\
\hline
&\multicolumn{2}{|c|}{$k=5$} &\multicolumn{2}{|c|}{$k=10$}&\multicolumn{2}{|c|}{$k=15$}&\multicolumn{2}{|c|}{$k=20$}&\multicolumn{2}{|c|}{$k=5$} &\multicolumn{2}{|c|}{$k=10$}&\multicolumn{2}{|c|}{$k=15$}&\multicolumn{2}{|c|}{$k=20$}&\multicolumn{2}{|c|}{$k=5$} &\multicolumn{2}{|c|}{$k=10$}&\multicolumn{2}{|c|}{$k=15$}&\multicolumn{2}{|c|}{$k=20$}  \\
\hline
&$\dag$&$\ddag$&$\dag$&$\ddag$&$\dag$&$\ddag$&$\dag$&$\ddag$&$\dag$&$\ddag$&$\dag$&$\ddag$&$\dag$&$\ddag$&$\dag$&$\ddag$&$\dag$&$\ddag$&$\dag$&$\ddag$&$\dag$&$\ddag$&$\dag$&$\ddag$\\
\hline
&&&&  &&&& &&&&&&&&  &&&& &&&&\\
$Dow$      &311 &151 &1000 &1000 &1000 &1000 &1000 &1000         &121 &46 &999 &991 &999 & 998&1000 &1000      &69 &22 &990 &968 &998 &997 &1000 &995 \\

$Nik$      &433 & 253&1000 &999 &1000 &1000 &1000 &1000        &176 &74 &993 &985 & 998&997 &1000 &999       &98 &36 &983 &963 &997&991 &999 &993 \\

 $UK$      &167 &77 &998 &995 &1000 & 999& 999& 998         &74 & 22&991 &976 &998 &997 &998 & 997       & 41&7 &982 & 943& 996&990 & 997&992 \\

 $AU$      &301 &142 &1000 & 999& 999&999 &1000 &1000            & 116&39 &997 &990 &998 &994 &1000 &999             &58 & 23& 990& 966&993 &989 &999 &995 \\

 $TB1$      &428 &227 &1000 &1000 &1000 & 999&999 &999            & 146& 55&993 & 976&997 &991 &998 & 996         & 75& 24& 976& 934&990 & 970&996 &989 \\

 $TB2$     &453 &256 & 999& 995&998 &997 &1000 &999          &159 &60 & 987&959 &994 & 982& 996&986            & 86& 21&958 & 899& 985& 961& 985&960 \\
&&&&  &&&& &&&&&&&&  &&&& &&&&\\
\hline

\hline
&\multicolumn{8}{|c|}{$\tau=25$}&\multicolumn{8}{|c|}{$\tau=50$}&\multicolumn{8}{|c|}{$\tau=100$}\\
\hline
&\multicolumn{2}{|c|}{$k=5$} &\multicolumn{2}{|c|}{$k=10$}&\multicolumn{2}{|c|}{$k=15$}&\multicolumn{2}{|c|}{$k=20$}&\multicolumn{2}{|c|}{$k=5$} &\multicolumn{2}{|c|}{$k=10$}&\multicolumn{2}{|c|}{$k=15$}&\multicolumn{2}{|c|}{$k=20$}&\multicolumn{2}{|c|}{$k=5$} &\multicolumn{2}{|c|}{$k=10$}&\multicolumn{2}{|c|}{$k=15$}&\multicolumn{2}{|c|}{$k=20$}  \\
\hline
&$\dag$&$\ddag$&$\dag$&$\ddag$&$\dag$&$\ddag$&$\dag$&$\ddag$&$\dag$&$\ddag$&$\dag$&$\ddag$&$\dag$&$\ddag$&$\dag$&$\ddag$&$\dag$&$\ddag$&$\dag$&$\ddag$&$\dag$&$\ddag$&$\dag$&$\ddag$\\
\hline
&&&&  &&&& &&&&&&&&  &&&& &&&&\\
$Dow$    & 24& 5&939 &858 &990 &964 &985 &966            &9 &3 & 807&677 &940 &887 &948& 872                    &4 &1 &566 & 381& 811& 669&808 &686 \\

$Nik$    & 34& 5& 920& 809& 982&848 &977 & 930         &11 & 2&764 &581 &914 &831 &897 &812                &4 & 1& 485& 281&750 &582 & 742& 575\\

 $UK$    & 11& 1 &929 &843&982 &942 &979 & 953           &4 &1 & 789& 630& 919& 840& 926& 843        & 1& 1&541 & 327&783 & 632&774 &640 \\

$AU$     & 23& 5&931 &860 & 983&949 & 983& 956             &6 & 2& 816& 666& 921&852 &931 &  846        & 4& 1& 561&353 &776 & 648& 786& 649\\

 $TB1$   &25 & 4&876 &765 &946 & 870&965 &893           &5 &1 & 698& 519& 822& 711&846 & 712           & 1& 1& 418& 230&627 & 415&604 &400 \\

 $TB2$   &21 & 6&844 &696 &933 & 851& 928& 859          & 10& 3& 627& 446& 798& 638& 807&  657            &3 & 1& 368&167 & 534& 312&544 &336 \\
&&&&  &&&& &&&&&&&&  &&&& &&&&\\

 \hline
\end{tabular}\\[0.5ex]
}\scriptsize{Note: $k=5$, $k=10$, $k=15$ and $k=20$ refer to the
number of rejections at 95\% ($\dag$) and 99\% ($\ddag$)
confidence levels (these intervals are given by [0.809, 1.862] and
[0.721, 2.098],
\\respectively) for the 1000 simulated time series.}
\end{table}

\end{landscape}


\begin{landscape}
\begin{table}[!h]
\caption{Lo's modified R/S Hurst exponent $H$ values for the
empirical and simulated data.}\label{Lo'sH}{\scriptsize
\begin{tabular}{|c|ccccc|ccccc|ccccc|}
\hline &\multicolumn{5}{|c|}{$\tau=0$}&\multicolumn{5}{|c|}{$\tau=5$}&\multicolumn{5}{|c|}{$\tau=10$}\\
\hline
& $Emp$ & $k=5$ & $k=10$ & $k=15$ & $k=20$ &$Emp$ & $k=5$ & $k=10$ & $k=15$ & $k=20$&$Emp$ & $k=5$ & $k=10$ & $k=15$ & $k=20$  \\
\hline
&&&&&  &&&&& &&&&&\\
$Dow$    & 0.620 &0.556 &\textbf{0.674} &\textbf{0.703} &\textbf{0.704}      &0.607 &0.540&\textbf{0.650} &\textbf{0.677} &\textbf{0.678}      & 0.597 &0.532&\textbf{0.636} &\textbf{0.662} &\textbf{0.663}                     \\
          &&(0.024)&(0.029)&(0.030)&(0.031)                  &&(0.024)&(0.028)&(0.029)&(0.030)                      &&(0.024)&(0.028)&(0.028)&(0.029)\\
$Nik$     & 0.723 &0.564 &0.670 &\textbf{0.695} &\textbf{0.695}              & 0.705&0.544 &0.643 &\textbf{0.667} &\textbf{0.667}             & 0.693 &0.535 &0.629 &\textbf{0.652} &\textbf{0.651}                    \\
          &&(0.025)&(0.027)&(0.028)&(0.029)                  &&(0.025)&(0.027)&(0.028)&(0.029)                      &&(0.025)&(0.027)&(0.027)&(0.028)\\
 $UK$     &0.712  &0.545 &\textbf{0.665}&\textbf{0.694} &\textbf{0.696}      &0.696 &0.532 &0.644 &\textbf{0.672} &\textbf{0.673}         &0.685 &0.525 &0.632 &\textbf{0.658} & \textbf{0.660}                    \\
          &&(0.025)&(0.030)&(0.033)&(0.036)                  &&(0.025)&(0.029)&(0.032)&(0.035)                      &&(0.025)&(0.029)&(0.031)&(0.034)\\
$AU$     &0.726  &0.555 &0.673 &\textbf{0.700} &\textbf{0.701}                       &0.711 &0.539 &0.650 &\textbf{0.674} &\textbf{0.676}       &0.700  &0.531 &0.636 &\textbf{0.660} &\textbf{0.661}                     \\
          &&(0.025)&(0.029)&(0.032)&(0.032)                  &&(0.025)&(0.028)&(0.031)&(0.031)                      &&(0.025)&(0.028)&(0.030)&(0.030)\\
 $TB1$    &0.746 &0.565 &0.670 &\textbf{0.689} &\textbf{0.691}               &0.721 &0.547 &0.642 &\textbf{0.660} &\textbf{0.661}            & 0.704 &0.535 &0.627 &\textbf{0.644} & \textbf{0.645}                    \\
          &&(0.024)&(0.028)&(0.031)&(0.029)                  &&(0.024)&(0.028)&(0.030)&(0.028)                      &&(0.024)&(0.028)&(0.029)&(0.028)\\
 $TB2$    & 0.724 &0.567 &0.662 &\textbf{0.679} & \textbf{0.680}            &0.704 &0.545 &0.634 &\textbf{0.650} &\textbf{0.652}              & 0.689 &0.536 &0.620 &\textbf{0.636} & \textbf{0.637}                   \\
          &&(0.025)&(0.028)&(0.028)&(0.028)                  &&(0.025)&(0.027)&(0.028)&(0.028)                      &&(0.024)&(0.027)&(0.028)&(0.027)\\

\hline
\hline &\multicolumn{5}{|c|}{$\tau=25$}&\multicolumn{5}{|c|}{$\tau=50$}&\multicolumn{5}{|c|}{$\tau=100$}\\
\hline
& $Emp$ & $k=5$ & $k=10$ & $k=15$ & $k=20$ &$Emp$ & $k=5$ & $k=10$ & $k=15$ & $k=20$&$Emp$ & $k=5$ & $k=10$ & $k=15$ & $k=20$  \\
\hline
&&&&&  &&&&& &&&&&\\
$Dow$     & 0.578 &0.521&\textbf{0.612} &\textbf{0.636} & \textbf{0.637}             &0.560 &0.513&\textbf{0.592} &\textbf{0.614} &\textbf{0.615}           & 0.542 &0.508&\textbf{0.571} &\textbf{0.590} &\textbf{0.591}                     \\
          &&(0.024)&(0.027)&(0.027)&(0.028)                  &&(0.023)&(0.026)&(0.026)&(0.027)                      &&(0.023)&(0.025)&(0.025)&(0.026)\\
$Nik$     & 0.671 &0.522 &0.605 &\textbf{0.627} &\textbf{0.626}    & 0.650&0.514 &0.586 &\textbf{0.606} &\textbf{0.605}       &0.628  &0.509 &0.566 &\textbf{0.584} &\textbf{0.583}                     \\
         &&(0.025)&(0.026)&(0.027)&(0.027)                  &&(0.024)&(0.026)&(0.026)&(0.026)                      &&(0.024)&(0.025)&(0.024)&(0.025)\\
 $UK$      & 0.662 &0.515 &0.610 &\textbf{0.634} & \textbf{0.635}       &0.641 &0.508 &0.590 &\textbf{0.612} & \textbf{0.613}           &0.617 &0.503 &0.569 &\textbf{0.589} & \textbf{0.589}                   \\
          &&(0.025)&(0.028)&(0.029)&(0.032)                  &&(0.024)&(0.027)&(0.028)&(0.030)                      &&(0.024)&(0.026)&(0.026)&(0.028)\\
 $AU$     &0.679 &0.520 &0.612 &\textbf{0.634} & \textbf{0.635}                     &0.657 &0.512 &0.592 &\textbf{0.612} &\textbf{0.613 }        &0.631  &0.507 &0.571 &\textbf{0.588} &\textbf{0.589}                     \\
          &&(0.025)&(0.027)&(0.029)&(0.029)                  &&(0.024)&(0.026)&(0.027)&(0.027)                      &&(0.023)&(0.025)&(0.026)&(0.026)\\
 $TB1$    & 0.672 &0.522 &0.603 &\textbf{0.619} & \textbf{0.621}                    &0.642 &0.514 &0.583 &\textbf{0.597} &\textbf{0.598}          & 0.610 &0.509 &0.563 &\textbf{0.575} &\textbf{0.576}                    \\
          &&(0.024)&(0.027)&(0.028)&(0.027)                  &&(0.024)&(0.026)&(0.027)&(0.026)                      &&(0.023)&(0.025)&(0.026)&(0.024)\\
 $TB2$    & 0.661 &0.520 &0.597 &\textbf{0.611} &\textbf{0.612}                     &0.633 &0.514 &0.578 &\textbf{0.591} &\textbf{0.592}                   & 0.604 &0.509 &\textbf{0.559} &\textbf{0.571} & \textbf{0.571}                   \\
          &&(0.024)&(0.027)&(0.027)&(0.027)                  &&(0.024)&(0.026)&(0.026)&(0.026)                      &&(0.023)&(0.025)&(0.025)&(0.024)\\

 \hline
\end{tabular}\\[0.5ex]
}\scriptsize{Note: $Emp$ stands for the empirical value of Lo's
Hurst exponent, $k=5$, $k=10$, $k=15$ and $k=20$ refer to the mean
and standard deviation of Lo's Hurst \\exponent based on the
corresponding 1000 simulated time series with different $k$.
Boldface numbers are those cases in which empirical $H$s fall into the\\
corresponding 2.5 to 97.5 percent quantiles of the 1000
simulation-based values of $H$. }
\end{table}
\end{landscape}

\end{document}